# Pacific Neutrinos: Towards a High Precision Measurement of CP-Violation ?

Claude Vallée

*CPPM, Aix-Marseille Université, CNRS/IN2P3*
*163 Avenue de Luminy, Case 902, F-13288 Marseille cedex 09, France*
*and*
*DESY, Notkestraße 85, D-22607, Hamburg, Germany*

ABSTRACT

The application of deep sea low energy neutrino detection techniques to long baseline neutrino physics is investigated, with a focus on a possible configuration based on a FNAL neutrino beam impinging a detector hosted by the NEPTUNE/OOI submarine observatories offshore of British Columbia.

## 1 Introduction

Following the measurement of the neutrino mixing angle $\theta_{13}$, the focus of neutrino oscillation physics has moved to the determination of the neutrino mass hierarchy and the investigation of CP violation. The main two next generation projects in preparation, DUNE [1] and HyperKamiokande [2], are still based on underground detectors of intrinsically limited target sizes. Their far detectors will accumulate at most a few thousand $\nu_e$ CC events over the full program life times despite the planned neutrino beam power upgrades. This will limit the precision on the CP-violating phase $\delta_{CP}$ to O(10º).

Deep sea detection technologies suitable for neutrino measurements in the few GeV domain, currently under final development by the KM3NeT/ORCA consortium [3], may provide a means to bypass this limitation. The deep sea offers a potentially unlimited clean and stable transparent detection medium located in the dark and protected from cosmic rays by the water column. Its exploitation is only limited by the availability of deep sea marine deployment tools and related technologies, and by the available funding for detector sensors. A configuration based on a neutrino beam from Protvino pointing to one of the KM3NeT sites has been proposed [4] to determine the neutrino mass hierarchy. The establishment of FNAL as a long term neutrino facility, together with the development of the permanent deep sea cabled observatories NEPTUNE [5] and OOI [6] offshore of Vancouver and Seattle, now offer an outstanding opportunity to implement a high statistics neutrino oscillation project in the future.

## 2 Pacific neutrinos

The Pacific neutrinos concept builds on the recent KM3NeT deep sea neutrino detector developments, on the existing NEPTUNE/OOI submarine observatories, and on the FNAL neutrino beam capabilities.

*2.1 Deep sea neutrino detection technology*

Deep sea neutrino detectors have been originally developed for high energy neutrino astronomy. The pioneering ANTARES telescope [7] offshore of Toulon, France, has successfully operated 12 detector lines with 900 optical modules since 2008, proving the reliability of deep sea technologies on time scales of a decade. Based on this success, the KM3NeT consortium has recently developed a second generation instrumentation [3] suitable for much larger detectors. The main ingredient is a Digital Optical Module ("DOM") equipped with 31 3" PMTs. The KM3NeT DOM is a fully integrated and modular device, acting as an independent Ethernet hub and gathering all functionalities of an ANTARES line storey with a similar light collection area. Another improvement is the simplified deployment of the KM3NeT lines, made easier by depositing them compacted on the sea floor and unfurling them from the bottom. Altogether KM3NeT succeeded to reduce the detector cost by a factor 3 compared to ANTARES, the dominant contribution being that of the DOMs with a unit cost of ~10 K$.

The KM3NeT instrumentation can be deployed either in a coarse configuration for high energy neutrino astronomy, or with a denser granularity for the planned determination of the neutrino mass hierarchy with atmospheric neutrinos (ORCA project). In the latter case, DOMs are spaced from each other by down to 6 m within a line, and lines are deposited every 20 m on the seafloor. For the ORCA measurement the relevant energy range of atmospheric neutrinos is 5-10 GeV. KM3NeT has studied the Cerenkov patterns of such events within the detector in detail, and shown that the ORCA granularity allows measuring the neutrino flavor and kinematics with sufficient precision for the neutrino mass hierarchy determination [3].

*2.2 The NEPTUNE/OOI Observatories*

In the past years, Canada and the US have deployed worldwide unique deep sea cabled observatories offshore of Vancouver (NEPTUNE [5]) and Seattle (OOI [6]) at a depth of 2500 m, similar to that of the ANTARES neutrino telescope. These observatories are originally devoted to environmental sciences including oceanology, biogeochemistry, biology and seismology. The high electrical power supply and the large data throughput of the deep sea cabled network provide unprecedented observational opportunities compared to autonomous sensors commonly used in these domains.

The cabled infrastructures of NEPTUNE/OOI and KM3NeT are based on similar technologies: electrical power is transported offshore after conversion to high voltages of O(10 kV), and data are transferred to shore at a high rate via optical fibres using dense wavelength multiplexing. The sensors are connected to intermediate nodes distributed on the network using standard deep sea industrial wet-mateable connectors. The NEPTUNE and OOI observatories provide a unique marine expertise on site and host outstanding marine operation tools such as cable ship and Remote Operated Vehicles. Those are pre-requisites for the long term deployment and maintenance of a large deep sea detector.

*2.3 The Pacific neutrinos configuration*

A baseline longer than 2500 km is necessary for the muon neutrino maximal first oscillation to correspond to an energy higher than 5 GeV, which is the suitable range for a deep sea detector "à la ORCA". This condition would be met by a beam from FNAL impinging the NEPTUNE/OOI observatory: the distance between Chicago and NEPTUNE/OOI is ~3100 km, corresponding to a maximal first oscillation at ~6.2 GeV. The overall configuration of the Pacific neutrinos project is sketched in figure 1.

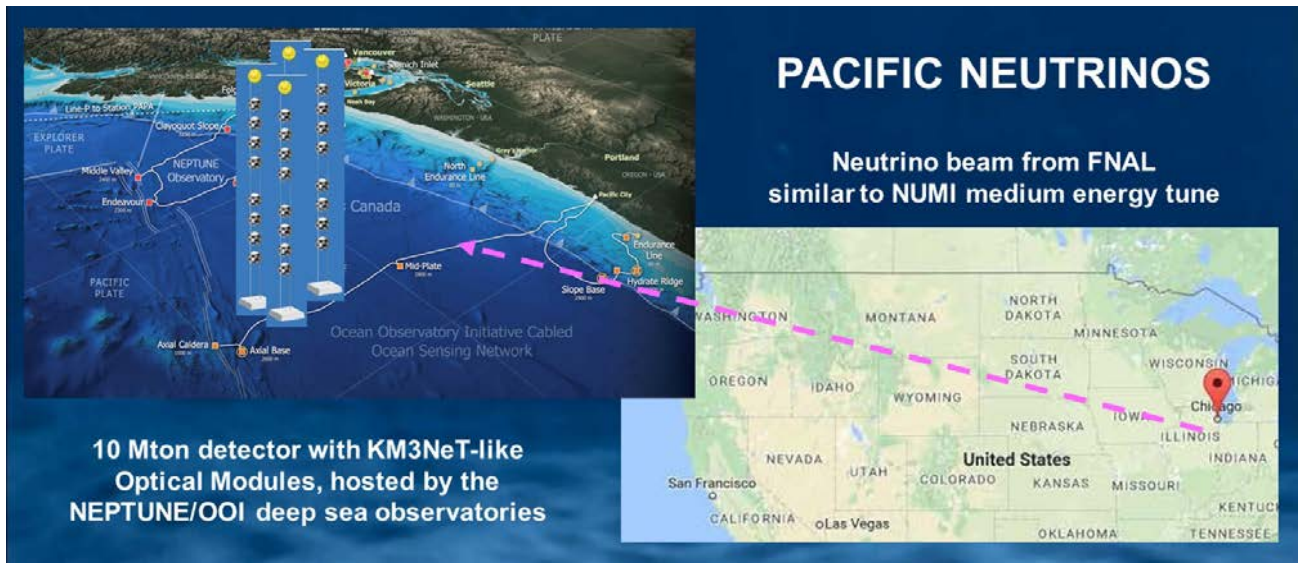

Figure 1: The Pacific neutrinos configuration. The baseline amounts to approximately 3100 km. The NEPTUNE/OOI observatories are deployed at a depth of 2500 m, similar to that of the ANTARES neutrino telescope.

The Pacific neutrinos beam horizontal direction is very close to that of the LBNF beam, but its vertical direction points 8° deeper due to the Earth curvature. The optimal beam spectrum is similar to that of the NUMI medium energy tune, seen by MINOS+ on the on-axis NOvA beam, and corresponding to a peak energy between 6 and 7 GeV. This energy range makes best use of the unique capabilities of the FNAL complex (with CERN) to provide high energy neutrino beams. It is also a domain where the neutrino cross sections and event topologies are under control, with measurements available from the past and new measurements easily feasible in the same energy range by e.g. the MINERvA experiment located on the on-axis NOvA beam. The beam power is not critical since a large statistics is achieved more economically from the large detector volume than by pushing beam technologies to their limits.

Assuming an instrumentation granularity of 1 DOM/kton water (corresponding to an inter-DOM distance of 2.5 m in detector lines distant by 20 m), a detector volume of O(10 Mton) could be instrumented at a cost of O(100 M$). With the current KM3NeT technology, such a detector would correspond to a few hundred instrumented lines and would require new cables and nodes to be added to the existing NEPTUNE/OOI network for its connection to the shore.

## 3 Potential precision

The statistical precision of the Pacific neutrinos configuration can be estimated from the LBNO study [8] performed with a baseline of 2300 km and a 24 kton LAr TPC detector. The exercise is done here for the showcase of a neutrino beam with the Normal Hierarchy hypothesis. Event numbers are extrapolated from those given in table 8 of ref. [8] using the following rescaling factors:
- Factor (10000 kton/24 kton) for the target mass.
- Factor (2300 km/3100 km)$^2$ for the lateral beam dispersion on the longer Pacific neutrinos baseline.
- Factor (3100 km/2300 km) for the neutrino interaction cross section proportional to its energy, which has an average value on the first oscillation peak proportional to the baseline.
- Factor (120 GeV/400 GeV) to normalize the number of events to a given number of Protons On Target (POT) of the FNAL Main Injector, taking into account its lower proton energy compared to the CERN SPS considered by LBNO.

This simple rescaling was checked to be compatible within 30% with an independent rescaling from the MINOS+ data collected in a beam similar to the Pacific neutrinos optimal beam. A more precise evaluation should further take into account the exact Pacific neutrinos beam optics, the specifications of the water target and the detector efficiency, which requires detailed simulations.

The resulting expected numbers of the various categories of Pacific neutrinos interactions are given in table 1 in the case of no CP violation $\delta_{CP} = 0°$, for a total number of $10^{21}$ POT roughly corresponding to one year of equivalent LBNF operation.

Table 1: Expected number of collected events extrapolated from the LBNO study (see text) for $10^{21}$ POT corresponding to ~1 year of equivalent LBNF operation, in the case of a neutrino beam with the Normal Hierarchy hypothesis and no CP violation. The number of NC events is estimated from the number of un-oscillated $\nu_\mu$ using the approximate relation $\sigma(NC) \sim 1/3\ \sigma(CC)$. The number of $\nu_\tau$ CC events is an underestimation since the kinematic suppression due to the $\tau$ mass is weaker for the Pacific neutrinos beam energy than for LBNO.

| PACIFIC NEUTRINOS Number of events | $\nu_\mu$ un-osc. CC | $\nu_\mu \to \nu_e$ CC | $\nu_\mu \to \nu_\mu$ CC | $\nu_\mu \to \nu_\tau$ CC | $\nu_\mu \to \nu_{all}$ NC |
|---|---|---|---|---|---|
| 10 Mton detector FNAL $\nu$ beam $10^{21}$ POT NH and $\delta_{CP} = 0°$ | 1 030 000 | 57 000 | 280 000 | > 60 000 | ~ 340 000 |

The Pacific neutrinos configuration has the potential to increase the number of collected events by two orders of magnitude compared to the DUNE and HyperKamiokande projects. The main signature for the measurement of the CP-violating phase is the $\nu_e$ CC appearance signal. The expected number of such events, given for $\delta_{CP} = 0°$ in table 1, varies by ± 30% on the $2\pi$ range of possible $\delta_{CP}$ values. The Pacific neutrinos have therefore the potential to measure $\delta_{CP}$ with a statistical precision of ~1°. The $\nu_\tau$ CC appearance sample is also of high interest to constrain the unitarity of the neutrinos mixing matrix. With an energy well above the production threshold of $\tau$ leptons, the Pacific neutrinos offer a unique opportunity to measure a high statistics sample of $\nu_\tau$ CC events and provide a strong constraint on the PMNS matrix. It is worth noting that the expected number of $\nu_\tau$ extrapolated from LBNO in table 1 is underestimated, since the lower energy LBNO configuration suffers from a larger kinematic suppression of $\tau$ production.

The actual precision reachable with Pacific neutrinos strongly depends on the ability to discriminate the relevant signal events from backgrounds. The various categories of events have very distinctive patterns in terms of particle identification and global event kinematic variables. For CP violation the typical S/B ratio is 0.1 (table 1). The $\nu_e$ CC signal events are characterized by a high energy isolated electron and a balanced total transverse momentum, whereas the associated backgrounds (NC events with a $\pi^0$, $\nu_\mu$ CC with a high energy bremsstrahlung photon from the muon, $\nu_\tau$ CC in the e decay channel) all have lower energy e candidates in average, and/or significant missing transverse momentum. Muons can be identified from the characteristic timing structure of the Cerenkov pattern of long penetrating tracks, and their energy estimated from their absorption range. The $\tau$ decay into a muon is a promising channel for a precise measurement of $\nu_\tau$ appearance.

A critical aspect of the Pacific neutrinos evaluation is to assess which detector granularity is needed to perform the event detection and reconstruction at the level required to fully benefit from the high statistical precision of the samples. To this respect it is important to note that the large statistics of the various event categories allow using statistical methods to estimate their populations, and that a perfect event by event discrimination is not needed. In particular, the requirements on $e/\pi^0$ separation are weaker than for HyperKamiokande or DUNE. On the other hand it is crucial to control the statistical separation with a high accuracy. To our knowledge, these are issues which have not yet been addressed in depth for beam neutrino large Cerenkov detectors in this energy range.

## 4 Detector layout and response

The KM3NeT/ORCA proponents have studied in detail [3] the Cerenkov light emission patterns of neutrino events in the 5-10 GeV domain. In water, Cerenkov light is emitted mostly on a cone of 43º along the particle direction. In the few GeV domain, light emission from electrons and hadronic showers is concentrated in one or two meters, whereas muons have a much longer mean emission path. The Cerenkov light ring patterns are conserved in the full light absorption length (~60 m) thanks to the high light scattering length (>100 m) in water. Electron Cerenkov rings have a distinctive high intensity whereas hadron rings are fainter and more difficult to detect.

The KM3NeT/ORCA baseline layout used in the LoI simulations [3] is adapted to detection of atmospheric neutrinos in the 5-10 GeV range. It consists of an array of 115 vertical detection lines with an interline spacing of 20 m. This interline distance is considered to be a minimum compatible for the precision of line positioning on the seabed and the highest deep Mediterranean currents. Each line hosts 18 DOMs spaced from each other by 6 m. Using an event reconstruction adapted to atmospheric neutrinos of initially unknown direction, the consortium has shown [3] that this configuration allows reconstruction of $\nu_e$ CC events with precisions of ~1 m on the vertex position, ~5º (resp ~20º) on the electron (resp. hadronic) shower direction, and ~20% on the total energy.

A similar detector layout located in a beam is expected to show significantly better performances thanks to several additional constraints. With a horizontal beam, the Cerenkov rings cross 3 successive transverse planes of detection lines in average on the 60 m water light absorption range. If the line positions are staggered from one plane to another in the horizontal direction perpendicular to the beam, the DOMs offer an effective transverse granularity of ~6 m in both transverse directions for reconstruction of Cerenkov rings. The neutrino beam pulse timing also strongly suppresses asynchronous background from downstream cosmic muons and bioluminescence, which should allow lower initial thresholds to be used for trigger and pattern reconstruction. The latter will also be simplified by the beam direction, which gives a natural timing ordering of expected neutrino signals and provides a fixed reference for the reconstruction of kinematic quantities like transverse momenta. Altogether these elements should greatly increase the detector efficiency at low energy and the resolution of reconstructed kinematic variables.

If needed an increased DOM granularity compared to ORCA can be achieved by reducing the inter-DOM vertical distance within a line. Mechanical structures other than independent vertical lines could also be envisaged to further increase the instrumentation granularity. It has however to be kept in mind that simple mechanical structures are essential for reliable deployment of deep sea components.

A quantitative evaluation of the Pacific neutrinos configuration has recently started. The study includes a detailed simulation of the response of 100 detection lines hosted on the Cascadia site of the NEPTUNE Observatory, in a neutrino beam from FNAL slightly adapted from the NUMI medium energy tune. First simulations of $\nu_e$ CC events in the detector confirm that the electron Cerenkov rings are highly visible. The simulations will be used to develop efficient methods for identification and reconstruction of the various event categories, and for a quantitative extraction of their relative populations.

Depending on the results, the next steps should include: a comprehensive optimization of the detector layout and beam configurations (ν and anti-ν share) for an optimal measurement of $\delta_{CP}$ and other oscillation parameters; further development and validation of the corresponding deep sea technologies within the ORCA project and/or through connection of prototypes to the existing NEPTUNE/OOI infrastructure; and detailed characterization of the properties of the deep sea site candidates to host a large detector. Within this program attention should be brought to the potential of a higher density instrumented core for measurements on the second oscillation peak around 2 GeV. It is also worth remembering that such a large high granularity deep sea detector would have a broader physics case including e.g. investigation of extra-terrestrial low energy neutrino signals.

## 5 Outlook

Deep sea instrumentation for low energy neutrino detection is reaching maturity and may provide the optimal compromise between detector size and instrumentation granularity for the long term future of long baseline neutrino physics. With the FNAL neutrino facility and the NEPTUNE/OOI deep sea cabled observatories, North America offers the geographical opportunity and the institutional synergies necessary for a concrete implementation of such a project. In the ongoing design of the FNAL neutrino facility sufficient flexibility should be maintained to facilitate longer term upgrades towards a next generation of projects.

**Acknowledgements**

I am grateful to Jürgen Brunner, Darren Grant, Elisa Resconi, Stefan Schönert, Jarred Barron, Jannik Hofestädt, Claudio Kopper and Chris Weaver for fruitful discussions and contributions to the definition of the Pacific neutrinos project.